\documentclass[apl,superscriptaddress,preprint,endfloats]{revtex4-1}

\newcommand {\SRO}{Sr$_{\mathrm{2}}$RuO$_{\mathrm{4}}$}

\usepackage{graphicx}
\usepackage{bm}
\usepackage{pifont}
\usepackage{amsmath}
\begin{document}
\title{Molecular beam epitaxy growth of superconducting {\SRO} films}
\author{M. Uchida}
\email[Author to whom correspondence should be addressed: ]{uchida@ap.t.u-tokyo.ac.jp}
\affiliation{Department of Applied Physics and Quantum-Phase Electronics Center (QPEC), University of Tokyo, Tokyo 113-8656, Japan}
\author{M. Ide}
\affiliation{Department of Applied Physics and Quantum-Phase Electronics Center (QPEC), University of Tokyo, Tokyo 113-8656, Japan}
\author{H. Watanabe}
\affiliation{Department of Applied Physics and Quantum-Phase Electronics Center (QPEC), University of Tokyo, Tokyo 113-8656, Japan}
\author{K. S. Takahashi}
\affiliation{RIKEN Center for Emergent Matter Science (CEMS), Wako 351-0198, Japan}
\affiliation{PRESTO, Japan Science and Technology Agency (JST), Chiyoda-ku, Tokyo 102-0075, Japan}
\author{Y. Tokura}
\affiliation{Department of Applied Physics and Quantum-Phase Electronics Center (QPEC), University of Tokyo, Tokyo 113-8656, Japan}
\affiliation{RIKEN Center for Emergent Matter Science (CEMS), Wako 351-0198, Japan}
\author{M. Kawasaki}
\affiliation{Department of Applied Physics and Quantum-Phase Electronics Center (QPEC), University of Tokyo, Tokyo 113-8656, Japan}
\affiliation{RIKEN Center for Emergent Matter Science (CEMS), Wako 351-0198, Japan}

\begin{abstract}
We report growth of superconducting {\SRO} films by oxide molecular beam epitaxy (MBE). Careful tuning of the Ru flux with an electron beam evaporator enables us to optimize growth conditions including the Ru/Sr flux ratio and also to investigate stoichiometry effects on the structural and transport properties. The highest onset transition temperature of about 1.1 K is observed for films grown in a slightly Ru-rich flux condition in order to suppress Ru deficiency. The realization of superconducting {\SRO} films via oxide MBE opens up a new route to study the unconventional superconductivity of this material.
\end{abstract}
\maketitle

The layered perovskite {\SRO} has attracted enduring interest since Y. Maeno {\it et al.} found superconductivity in its single-crystalline bulk \cite{SRO}. Its fascinating properties as a possible two-dimensional chiral $p$-wave superconductor, classified into a topological superconductor, have been intensively studied from both the experimental and theoretical sides \cite{SROsymmetry, SROreview1, reviewdesiringfilm1, reviewdesiringfilm2}. In spite of the lasting experimental progress as represented by strain effects \cite{strain1, strain2}, its underlying physics has not been entirely understood. In this context, reproducible growth of superconducting thin films has long been desired in order to enable junction and microfabricated device experiments for determining pairing symmetry and topological aspects of the superconductivity \cite{reviewdesiringfilm1, reviewdesiringfilm2}.

Growth of superconducting {\SRO} thin films is known to be extremely difficult, because the low transition temperature ($T_{\mathrm{c}}\sim1.5$ K) in {\SRO} bulks is highly sensitive to impurities \cite{sensitivity} and sample nonstoichiometry \cite{SROsymmetry}. Among the many {\SRO} films grown by the pulsed laser deposition (PLD) method \cite{YoshiharuPLD, RobinsonPLD, antiphaseboundary1PLD, antiphaseboundary2PLD, otherSROfilm1PLD, otherSROfilm2PLD, otherSROfilm3PLD, otherSROfilm4PLDlaser, otherSROfilm5PLD, otherSROfilm6PLD, otherSROfilm8PLD}, successful growth of superconducting films have been very limited \cite{YoshiharuPLD, RobinsonPLD}. An alternative growth method is molecular beam epitaxy (MBE), which has traditionally delivered high-quality and high-reproducibility thin films in the field of semiconductors, but has now been adapted for the growth of oxides \cite{SchlomMBEreview}. In particular, it has found success in the growth of clean systems, such as those which display high mobility or unconventional superconductivity \cite {oxideMBEpower1, oxideMBEpower2, oxideMBEpower3, oxideMBEpower4}. Nonetheless, MBE growth of superconducting {\SRO} films has been highly challenging in spite of recent developments \cite{otherSROfilm7MBE, SchlomSROMBE, StemmerSROMBE, WOE}. The primary challenge in the MBE growth is to evaporate high-purity Ru while maintaining its stable flux through film deposition.

Here we demonstrate the growth of superconducting {\SRO} films using MBE with an electron beam evaporator. Careful tuning of the Ru flux enables us to perform systematic optimization of growth conditions to realize superconducting {\SRO} films.

\begin{figure}
\begin{center}
\includegraphics*[width=13.5cm]{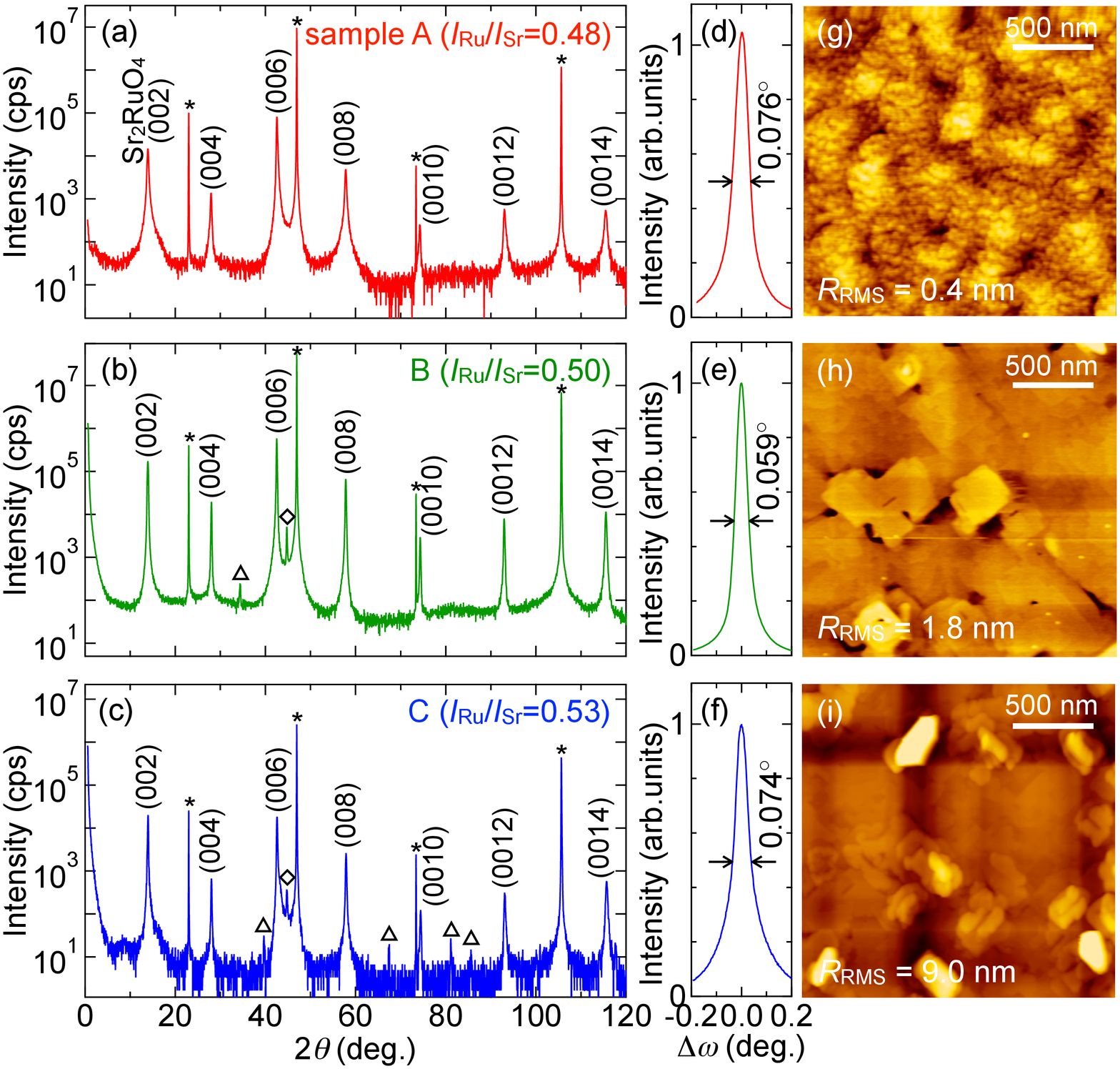}
\caption{
(a)--(c) XRD $\theta$--2$\theta$ scans, (d)--(f) rocking curves of the {\SRO} (006) peak, and (g)--(i) AFM images, for the samples A--C grown with different ratios between the Ru and Sr fluxes. LSAT substrate peaks in the XRD scans are marked with an asterisk. Tiny peaks denoted by a triangle or a diamond are respectively ascribed to RuO$_2$ or other Ruddlesden-Popper phase.
}
\label{fig1}
\end{center}
\end{figure}

The $c$-axis oriented {\SRO} films were grown with a Veeco GEN10 oxide MBE system on as-received single crystalline (001) (LaAlO$_{3}$)$_{0.3}$(SrAl$_{0.5}$Ta$_{0.5}$O$_{3}$)$_{0.7}$ (LSAT) substrates supplied by Furuuchi Chemical Co. 4N Sr and 3N Ru elemental fluxes were simultaneously provided from a conventional Knudsen cell and a Telemark TT-6 electron beam evaporator, respectively. While the Sr flux $I_{\mathrm{Sr}}$, measured by an INFICON quartz crystal microbalance system, was set to $6.9\times10^{13}$ $\mathrm{atoms}/\mathrm{cm}^2 \mathrm{s}$, the Ru flux $I_{\mathrm{Ru}}$ was tuned to $3.3$, $3.4$, and $3.6\times10^{13}$ $\mathrm{atoms}/\mathrm{cm}^2 \mathrm{s}$ for samples A, B, and C, which correspond to $I_{\mathrm{Ru}}/I_{\mathrm{Sr}}=$ 0.48 (Ru-deficient), 0.50 (stoichiometric), and 0.53 (Ru-rich). Superconducting {\SRO} films were not grown out of this flux ratio range ($I_{\mathrm{Ru}}/I_{\mathrm{Sr}}=$ 0.48--0.53). Other conditions were the same for these three samples. The deposition was performed in 100\% $\mathrm{O}_{3}$ with a pressure of $1\times 10^{-6}$ Torr, supplied from a Meidensha Co. MPOG-104A1-R pure ozone generator, and at a substrate temperature of 900 $^{\circ}$C, achieved with a semiconductor-laser heating system \cite{otherSROfilm4PLDlaser}. The film thickness was about 58 nm and the growth rate was about 1.4 nm/min.

Figure 1 summarizes structural characterization of the three samples A--C grown with the different Ru/Sr flux ratios. As seen in x-ray diffraction (XRD) $\theta$--2$\theta$ scans (Figs. 1(a)--(c)), sharp (00$l$) {\SRO} peaks ($l$: even integer) are commonly observed up to the (0014) peak, indicating $c$-axis oriented epitaxial film growth. Tiny peaks assigned to Ru-rich phases such as RuO$_2$ or other Ruddlesden-Popper phases \cite{RP1, RP2} appear for samples B and C, while no impurity peaks are confirmed for sample A. From a thermodynamical standpoint, this result can be conversely interpreted as sample A having a non-negligible amount of Ru deficiency. In fact, sample B shows the sharpest film rocking curve among them in spite of the impurity peaks (Figs. 1(d)--(f)), although all the three values of the full width at half maximum (FWHM) are small enough to demonstrate high-quality oxide MBE growth. While the $a$-axis lattice constant is fixed to 3.87 {\AA} ($-0.07$\% compared to the bulk value \cite{latticeconstant}) on the LSAT substrate as confirmed in the reciprocal space mapping (not shown), the $c$-axis lattice constant estimated from the $\theta$--2$\theta$ scans is slightly elongated to 12.76 {\AA} ($+0.17$\%) for the three samples.

Surface topography taken by atomic force microscopy (AFM) (Figs. 1(g)--(i)) also consistently indicates changes reflecting the used Ru/Sr flux ratio. An extremely flat surface is confirmed for sample A. With increasing the Ru flux, on the other hand, ridge structures begin to be seen in sample B, and then some segregations presumably ascribed to RuO$_2$ appear on the surface of sample C. Accordingly, root mean square roughness $R_{\mathrm{RMS}}$ becomes much larger.

\begin{figure}
\begin{center}
\includegraphics*[width=13.5cm]{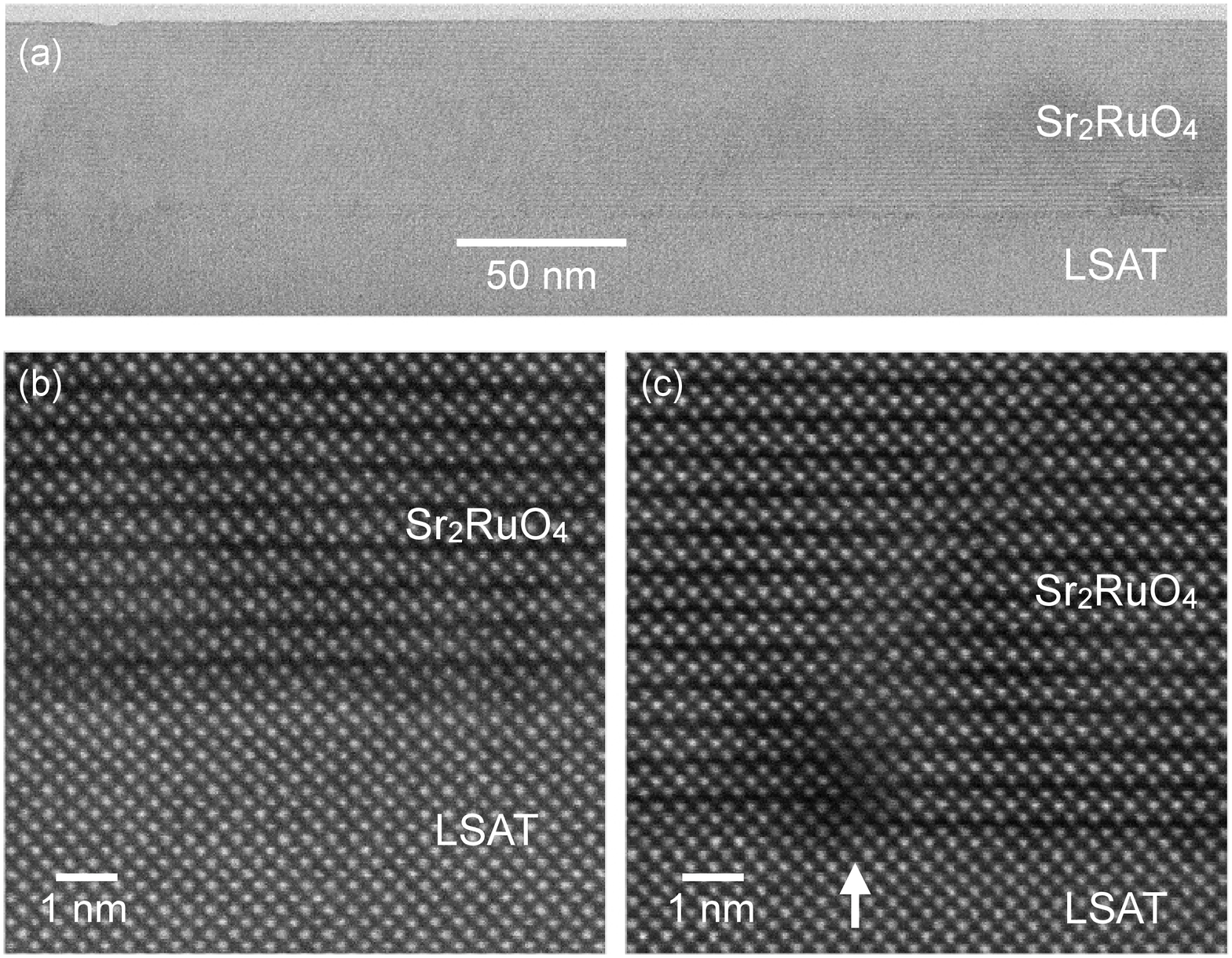}
\caption{
(a) Cross-sectional TEM image of the sample C, showing no secondary phase segregations, stacking faults, nor extended defects including out-of-phase boundary in a wide film region, which is quite a contrast to previously reported {\SRO} films grown by PLD \cite{YoshiharuPLD, antiphaseboundary1PLD, antiphaseboundary2PLD}. Higher-resolution HAADF-STEM images showing (b) epitaxially connected lattice structures and (c) an out-of-phase boundary at the interface between the film and substrate.
}
\label{fig2}
\end{center}
\end{figure}

Figure 2 shows cross-sectional transmission electron microscope (TEM) images of sample C. As shown in a low magnification image (Fig. 2(a)), there can be seen almost no segregations, stacking faults, nor extended defects inside the wide film region, even while the secondary phases due to excess supply of Ru are detected in XRD and AFM as noted above. In particular, extended defects characteristic to layered perovskite oxides, out-of-phase boundaries \cite{antiphaseboundary1PLD, antiphaseboundary2PLD, OPBs}, are almost entirely eliminated, in stark contrast to the superconducting {\SRO} film grown by PLD \cite{YoshiharuPLD}. Lattice structures of the {\SRO} film and the LSAT substrate and their epitaxial relation can be clearly confirmed in the magnified image taken by high angle annular dark field (HAADF) scanning transmission electron microscopy (STEM) (Fig. 2(b)). By searching in wider regions, an out-of-phase boundary is detected as indicated by an arrow (Fig. 2(c)), which may intrinsically originate from the 3.87 {\AA}-high unit-cell step of the LSAT substrate. While the out-of-phase boundaries have been considered to strongly suppress the superconductivity in {\SRO} films \cite {antiphaseboundary1PLD, YoshiharuPLD}, the defect-free regions in the grown films exist over about 500 nm, which is an order of magnitude longer than the in-plane superconducting coherence length of 66 nm \cite{coherencelength}. 

\begin{figure}
\begin{center}
\includegraphics*[width=13.5cm]{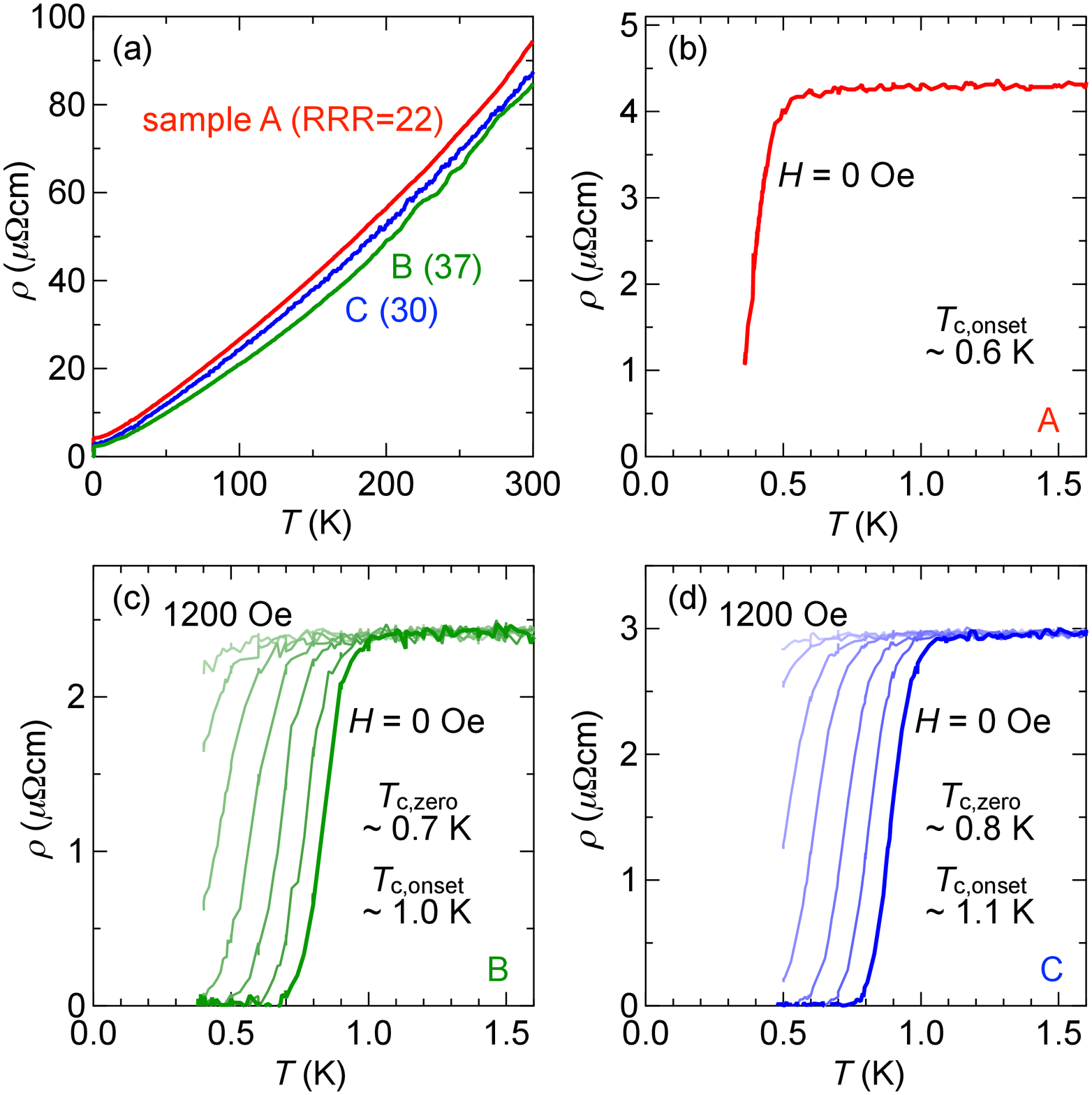}
\caption{
(a) Resistivity of the samples A--C as a function of temperature up to 300 K. (b)--(d) Low-temperature resistivity of the respective samples, measured with applying a magnetic field parallel to the $c$-axis at intervals of 200 Oe.
}
\label{fig3}
\end{center}
\end{figure}

Figure 3 summarizes the transport characteristics of the three samples A--C. Longitudinal resistivity was measured with a standard four-probe method in a Quantum Design PPMS cryostat equipped with a 9 T superconducting magnet and a 3He refrigerator. The samples show residual resistivity ratio (RRR = $\rho_{\mathrm{300 K}}/\rho_{\mathrm{2 K}}$) as high as 22, 37, and 30 (Fig. 3(a)). A clear superconducting transition is observed for all the samples, which is systematically suppressed with applying a magnetic field parallel to the $c$-axis. The highest transition temperature for sample C is $T_{\mathrm{c,zero}}\sim0.8$ K (zero resistivity) and $T_{\mathrm{c,onset}}\sim1.1$ K (onset), which exceeds $T_{\mathrm{c,zero}}\sim0.5$ K and $T_{\mathrm{c,onset}}\sim0.9$ K of previously reported superconducting films grown by PLD \cite{YoshiharuPLD}. Here it is notable that sample A shows a lower $T_{\mathrm{c}}$ than the other two samples having the surface segregations, indicating that the suppression of Ru deficiency in films is crucially important for realizing better superconductivity as also pointed out in the bulk experiments \cite{SROsymmetry}.

In conclusion, we have found a set of conditions to grow superconducting {\SRO} films in MBE. By carefully tuning the Ru flux supplied from the electron beam evaporator, we have systematically investigated the relationship between film stoichiometry and structure and transport properties. The highest transition temperature $T_{\mathrm{c,zero}}\sim0.8$ K and $T_{\mathrm{c,onset}}\sim1.1$ K is observed for the film grown in the slightly Ru-rich flux condition. Although segregations due to the excess supply of Ru are confirmed on the surface, it is concluded that the suppression of the Ru deficiency in the film is crucially important. Further precise control of the flux ratio as well as optimization of other growth conditions will be necessary to increase $T_{\mathrm{c}}$ to the bulk level ($T_{\mathrm{c}}\sim1.5$ K). The ability to grow high-quality superconducting {\SRO} films using oxide MBE opens a new avenue for verifying pairing symmetry and topological aspects of its superconductivity, for example through phase-sensitive transport measurements of mesoscopic systems \cite{mesoscopic1} and Josephson junctions \cite{josephson1, josephson2}.

This work was supported by and by a Grant-in-Aid for Scientific Research on Innovative Areas ``Topological Materials Science" No. JP16H00980 from MEXT, Japan and JST CREST Grant No. JPMJCR16F1, Japan. We thank J. Falson for proofreading of the manuscript.

\newpage
\end{document}